\documentclass[12pt]{article}
\usepackage{amsmath,amsfonts,amssymb,amsthm}
\usepackage{xcolor}
\usepackage{enumitem}
\usepackage{graphicx}
\usepackage[toc, page]{appendix}
\setitemize{itemsep=0.1pt}
\setenumerate{itemsep=0.1pt}
\RequirePackage[colorlinks,allcolors=blue]{hyperref}
\usepackage[affil-it]{authblk}
\usepackage{soul,xcolor}
\setul{}{1.2pt}
\usepackage[normalem]{ulem}
\setstcolor{red}

\begin{document} 

\renewcommand{\thefootnote}{\fnsymbol{footnote}}
\author[1,2]{Joel Steinegger}
\author[3]{Paul J. Steinhardt\thanks{Corresponding author: steinh@princeton.edu}}
\author[1,2]{Christoph Räth}
\author[4,3,5,6]{Salvatore Torquato}
\author[7,1,2]{Michael A. Klatt}

\affil[1]{\small German Aerospace Center (DLR), Institute of Frontier Materials on Earth and in Space, Functional, Granular, and Composite Materials, 51170 Cologne, Germany}
\affil[2]{\small Department of Physics, Ludwig-Maximilians-Universität München,\newline Schellingstr. 4, 80799 Munich, Germany}
\affil[3]{\small Department of Physics, Princeton University, Princeton, New Jersey 08544, USA}
\affil[4]{\small Department of Chemistry, Princeton University, Princeton, New Jersey 08544, USA}
\affil[5]{\small Princeton Materials Institute, Princeton University, Princeton, New Jersey 08544, USA}
\affil[6]{\small Program in Applied and Computational Mathematics, Princeton University,\newline Princeton, New Jersey 08544,USA}
\affil[7]{\small German Aerospace Center (DLR), Institute for AI Safety and Security,\newline Wilhelm-Runge-Str. 10, 89081 Ulm,Germany}

\title{Towards stealthy hyperuniform networks with optimal isotropic
complete photonic band gaps using a novel inverse design procedure} 

\date{\today}
\maketitle

\begin{abstract}
We present a two-stage inverse design procedure for producing
disordered stealthy hyperuniform trivalent photonic networks in two
dimensions with isotropic complete photonic band gaps (PBGs)  blocking
light regardless of direction or polarization (TE or TM) over a wide
frequency range. Most ordinary disordered systems fail to maintain
complete PBGs as system size increases. The only known exceptions that
remain open in the largest simulations have been generated by
mapping stealthy hyperuniform point patterns into trivalent networks.
However, the resulting networks are not truly stealthy hyperuniform
two-phase media. Although their PBGs remain open, they are
relatively narrow due to limited overlap between the TE and TM band
gaps and broad band tails caused by localized defect states. By
contrast, our two-stage inverse design aims to make the final network
itself stealthy hyperuniform, achieving unprecedented near-optimal overlap between
the TE and TM band gaps and a small defect state density at the band
edges. We obtain not only single realizations with large PBGs, but a
striking homogeneity across a large ensemble, effectively probing a
network with 100,000 vertices. This ensemble-based band gap is
comparable in width to the complete PBG of an anisotropic honeycomb
photonic crystal with the same network parameters and nearly an order
of magnitude wider than the previously widest known isotropic complete
PBGs. Our designs can be fabricated using additive manufacturing,
offering new pathways to manipulate electromagnetic waves for photonic
technologies.
\end{abstract}


%\linenumbers
\section{Introduction}\label{sec:intro}

A complete photonic band gap (PBG) is a spectral range of frequencies
within which the propagation of electromagnetic waves is forbidden for
all directions of incidence and both polarizations of the
waves~\cite{john_strong_1987, yablonovitch_inhibited_1987}. It is a
well-established phenomenon for certain crystalline
materials~\cite{joannopoulos_photonic_2008-1, ho_existence_1990-1,
maldovan_diamond-structured_2004, fu_connected_2005}. In such a
photonic crystal, the PBG arises from the periodic modulation of the
dielectric constant, and its formation can be understood and predicted
from a single unit cell via Bloch's theorem. On the other hand, the
theorem restricts the waves to conform to the crystal's periodicity
and hence to specific symmetries. In contrast, disordered solids can
be fully isotropic and could thus provide band gaps and
omnidirectional control of light that are independent of the incident
direction~\cite{florescu_designer_2009, wiersma_disordered_2013,
man_isotropic_2013, gorsky_engineered_2019, piechulla_tailored_2021,
tavakoli_over_2022, wan_hyperuniform_2023, gallego_hole-based_2025}.
However, whether such truly isotropic complete PBGs actually exist in
the thermodynamic limit remains a fundamental open
problem~\cite{torquato_hyperuniform_2018, yu_engineered_2021,
klatt_wave_2022, karcher_effect_2024}. 

Due to the absence of periodic order, Bloch's theorem does not apply in
disordered systems, and establishing the persistence of band gaps for
large system sizes requires computationally intensive simulations of
large samples and ensembles with high statistics. Assuming a
self-averaging property of the photonic density of states, such
an ensemble approach can probe larger system sizes that are otherwise
inaccessible  numerically for any single
realization~\cite{klatt_wave_2022}.
Such an ensemble analysis~\cite{klatt_wave_2022} revealed that complete
PBGs are not typical in two dimensional isotropic disordered systems
because most systems tend to lose their complete photonic band gap in
the large system limit. It was thus established that the only
remaining candidates are specially
designed isotropic, disordered trivalent
networks~\cite{florescu_designer_2009, klatt_wave_2022}. These networks
are derived from stealthy hyperuniform point
patterns~\cite{uche_constraints_2004, batten_classical_2008, torquato_ensemble_2015} that are
defined by a vanishing of the structure factor $S(k)=0$ for a range of
wavenumbers $0<k<K$. These point patterns were then mapped onto
trivalent networks by connecting the centroids of the Delaunay
triangulations and decorating the bonds and vertices with walls and
discs of a high-dielectric material~\cite{florescu_designer_2009}.
Remarkably, the complete PBG of the resulting networks remained open
even for the largest available simulation study~\cite{klatt_wave_2022}. 

\begin{figure}[tbp]
\centering
\includegraphics[width=\columnwidth]{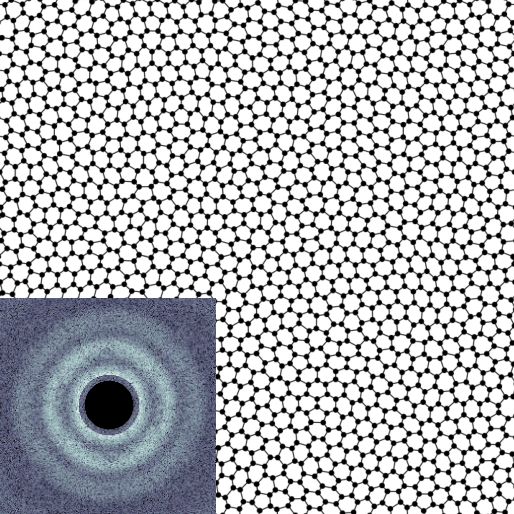}
\caption{Realization of a stealthy hyperuniform disordered network: A
representative trivalent network generated via our new inverse design
procedure ($\chi_1(K_1) = 0.37$, $\chi_2(K_2) = 0.24$), comprising 2,000
vertices. The main panel shows the pixelated grid used for band
structure computations in MPB. The inset displays the corresponding
spectral density $\tilde{\chi}_{\scriptscriptstyle V}(k)$ on a
logarithmic scale, confirming the isotropy and the stealthy hyperuniform
nature of the heterostructure.}
\label{fig:network_sample}
\end{figure}

While for these networks PBGs with gap-to-midgap ratios of up to
$\mathcal{O}(10\%)$ can be obtained for the best single realizations
with a few hundred vertices, there are strong fluctuations between small
realizations and the gaps decrease with system size down to
$\mathcal{O}(1\%)$ for the largest accessible system
sizes~\cite{klatt_wave_2022}. These limitations arise mainly from a
small spectral overlap between the band gaps for the transverse-electric
(TE) and transverse-magnetic (TM) polarizations, and also from a still
relatively high density of localized defect states within the gap that
give rise to broad and relatively shallow band tails. 

To overcome these limitations, we introduce a two-stage inverse design
method that produces networks that have not only wider PBGs at small
system sizes but also retain large isotropic and complete PBGs
comparable to photonic crystals even at the largest accessible system
sizes with 2000 vertices per realization. Furthermore, we are able to
generate many realizations (500) and use the ensemble
approach~\cite{klatt_wave_2022} that enables us to test for effective
system sizes of 100,000 vertices. The corresponding PBGs do not
significantly change compared to the PBGs for single realizations.

Notably, the previous networks are not truly stealthy hyperuniform
two-phase media. The high-fidelity stealthy hyperuniformity of the
progenitor point pattern is lost by the mapping that transforms the
stealthy hyperuniform point pattern to a trivalent network, as measured
by the spectral density $\tilde{\chi}_{\scriptscriptstyle
V}(k)$~\cite{torquato_random_2002}, which should vanish for $0<k<K$
in a stealthy hyperuniform two-phase
medium~\cite{torquato_hyperuniformity_2016}. The networks still
exhibit a strong suppression of single scattering for small wavenumbers,
but $\tilde{\chi}_{\scriptscriptstyle V}(k)$ is far from stealthy
hyperuniform for intermediate wavenumbers. These findings naturally
motivate the idea that a truly stealthy hyperuniform trivalent photonic
network will result in significantly large, complete PBGs in amorphous
photonic materials.

In contrast to the previous method, our new two-stage design method
directly targets stealthy hyperuniformity of the trivalent network
composed  of disks at the vertices and thin rectangular walls along all the
bonds, both consisting of the same high-dielectric material, rather than
the progenitor point pattern. The objective is to now obtain
high-fidelity stealthy hyperuniformity for the dielectric material
itself. We thus obtain a new class of isotropic, disordered
two-dimensional (2D) systems, see Fig.~\ref{fig:network_sample}. The single
scattering of these new heterostructures, as measured by the spectral
density, is suppressed by many orders of magnitude for wavenumbers
between zero and a finite cutoff $K$.

Our new procedure is necessary because first attempts with a naive
optimization of the previous networks that shifts the vertices to obtain
stealthy hyperuniformity for the heterostructures strongly distorts the
bond lengths and angles resulting in a closing of the band gaps. The
two-stage procedure introduced here bypasses this problem by allowing the
global network topology to change during the optimization, including the
simultaneous creation and deletion of vertices and the accompanying
rearrangement of bonds and rings. The new networks are composed of many
hexagonal cells, fewer five- and seven-sided cells, and only a few
eight-sided defect cells. Importantly, we observe almost no four-sided
cells (about one in 64000 cells), which are heuristically known to
introduce strong localized defect states in the band gap. All of our
cells are relatively regular with uniform bond angle and bond length
distributions.

\begin{figure}[tbp]
\centering
\includegraphics[width=0.96\columnwidth]{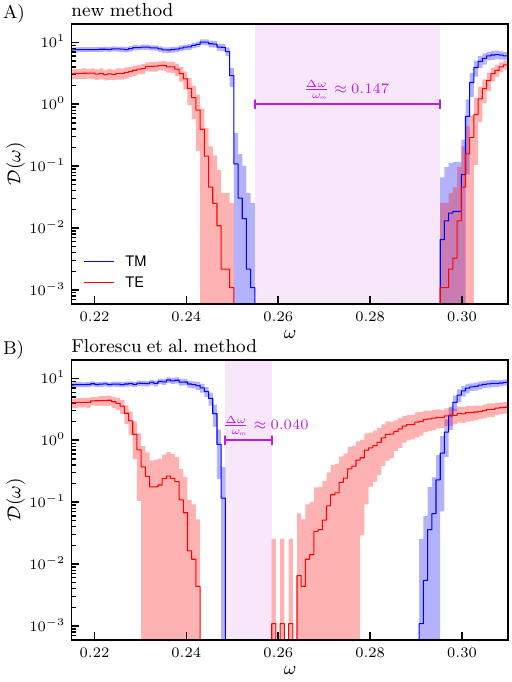}
\caption{Density of states $\mathcal{D}(\omega)$ averaged over 500
independent realizations for: A) the new inverse design
method ($r = 0.24$, $w = 0.08$), and B) the original design
approach by Florescu \textit{et al.}~\cite{florescu_designer_2009} ($r = 0.24$,
$w = 0.10$). Shaded regions represent the standard deviation within each
bin (with width $\Delta \omega \approx 0.00092$).
The dielectric contrast was $13.0$ in both cases, in alignment with
established values from earlier studies. 
}
\label{fig:DOS_comparison}
\end{figure}

The average density of states of an ensemble of these new structures
reveals that they exhibit nearly complete TE--TM band gap overlap and a
remarkably suppressed density of defect states within the band gap,
bordered by extremely steep band edges. The resulting isotropic complete PBGs are
substantially larger than what was previously possible; see
Fig.~\ref{fig:DOS_comparison}. The gap--to--mid-gap ratio of our
ensemble of isotropic, disordered networks (14.7\%) is comparable to
that of a honeycomb-based photonic crystal (16.7\%) with the same
dielectric contrast of 13, wall width, and disc radii.
This photonic crystal exhibits the largest known complete (though
anisotropic) PBG in two dimensions~\cite{fu_connected_2005}.

For our new stealthy hyperuniform heterostructures, we choose the cutoff
$K$ just below a six-fold orientational disorder-to-order phase
transition, to maximize structural correlations while maintaining a
fully amorphous, isotropic state. 
Although we optimize only for stealthy hyperuniformity with
a fixed local topology (i.e., with trivalent vertices), the structures
converge to networks that locally resemble the anisotropic
honeycomb-based photonic crystal 
even though our networks are isotropic.
Both the honeycomb-based crystal and our disordered networks completely
suppress single scattering over a broad range of wavenumbers near the
origin, i.e., they are both stealthy hyperuniform. The difference is,
while the ordered honeycomb lattice achieves this suppression up to the
first Bragg peak, our disordered trivalent networks exhibit a more
restricted range of wavenumbers over which single scattering is
suppressed within the $k$-space region. 

Our findings provide strong numerical support that the formation of
complete PBGs in isotropic, amorphous 2D systems is correlated with
high-$\chi$ stealthy hyperuniformity, as well as a trivalent topology
and favorable bond length and angle distributions.
It confirms the exceptional properties observed for stealthy
hyperuniform systems~\cite{torquato_ensemble_2015, zhang_ground_2015-1,
zhang_ground_2015}; these include besides band
gaps~\cite{florescu_designer_2009, froufe-perez_band_2017,
yu_engineered_2021, klatt_wave_2022, froufe-perez_bandgap_2023,
vynck_light_2023, siedentop_stealthy_2024, koga_stealthy_2026},
virtually optimal effective material
characteristics~\cite{torquato_multifunctional_2018,
kim_effective_2023}, transport
properties~\cite{leseur_high-density_2016,
gkantzounis_hyperuniform_2017, aubry_experimental_2020,
torquato_diffusion_2021, kim_unidirectional_2023,
diego_hypersonic_2025}, and (de-)localization
effects~\cite{park_deep-subwavelength_2025, meek_electromagnetic_2026,
barsukova_stealthy-hyperuniform_2026, klatt_transparency_2025,
vanoni_effective_2026}. Beyond their fundamental significance, these
results hold substantial promise for real-world applications, as the
emergence of high-precision additive manufacturing now permits the
fabrication of such complex structures across a broad range of length
scales.

\section{Design Protocol}

To date, the largest complete PBGs in disordered 2D
systems~\cite{florescu_designer_2009} that survive in the large system
limit~\cite{klatt_wave_2022} have been derived from stealthy
hyperuniform point patterns (with $\chi \approx 0.48$) and by applying a
Delaunay-centroidal algorithm from \cite{florescu_designer_2009} that
maps the point pattern to a trivalent network. The stealthy hyperuniform
progenitor point patterns have been obtained using the collective coordinate
procedure~\cite{uche_constraints_2004, batten_classical_2008,
torquato_ensemble_2015}. The degree of stealthy hyperuniformity is
characterized by the $\chi$-value of this point pattern. This value is
proportional to the ratio of the reciprocal space volume of wave vectors
with constrained values to the total number of degrees of freedom. In
two dimensions, $\chi(K)=K^{2}/(16\,\pi\,\rho)$, where $\rho$ is the
number density.

%with twice the number of vertices as points in the pattern.
This network is then transformed into a two-phase medium by placing
circular disks at the vertices of the trivalent network and connecting them with thin walls along
the edges, both consisting of the same material with a high dielectric
constant. The network inherits some beneficial features
from the initial stealthy hyperuniform pattern, including advantageous
physical properties~\cite{torquato_hyperuniform_2018} due in part to
their bounded hole size~\cite{zhang_can_2017, ghosh_generalized_2018},
which has been rigorously shown to be preserved by the
mapping~\cite{klatt_wave_2022}. However, the resulting heterostructure
is no longer stealthy hyperuniform: while the spectral density remains
suppressed at small wave vectors, this suppression vanishes at
intermediate $k$.

Here, we introduce a novel two-stage optimization procedure that targets
stealthy hyperuniformity of the two-phase medium directly. A brief
overview of the protocol is provided here; for a detailed description
see the SI.

In the first stage, dubbed the global topology optimization stage, we transform 
the conventional Delaunay-centroidal mapping algorithm into a targeted inverse 
design procedure.
We optimize a 2D point pattern such that the centroids of its Delaunay
triangulation become close to stealthy hyperuniform.
Hence, the loss function is directly associated with the centroids and not the
progentior points, in contrast to \cite{florescu_designer_2009}.
The stage proceeds iteratively: first, the Delaunay triangulation of the
progenitor point pattern and the centroids of the resulting triangles
are computed. Then, the structure factor $S(k)$ of the centroids is
evaluated for wave vectors $k \leq K_1$, and the loss function is
defined as the sum of $S(k)/k$ over this range (as in the collective
coordinate approach~\cite{uche_constraints_2004, batten_classical_2008,
torquato_ensemble_2015}). 
Finally, using a gradient-descent update, the points of the progenitor
pattern are displaced to minimize this loss function.
After each update, the Delaunay triangulation is recomputed based on the
new progenitor pattern, which will lead to changes in the network topology.
This cycle of triangulation, loss evaluation, and progenitor update is repeated
until the loss converges to a steady state, despite residual oscillations.

In the second stage, dubbed the network refinement stage, the centroid
pattern that achieved the lowest loss in the first stage is selected. It
defines the global topology of the trivalent network: network vertices
correspond to triangle centroids, and edges are formed between vertices
whose associated triangles share a common edge. The network is decorated
with discs and walls and subsequently optimized for stealthy
hyperuniformity. More precisely, the decorated network is modeled as a
two-phase system: circular disks are placed at the vertices, and thin
rectangular walls are added along the edges, both composed of the same
high-dielectric material. The vertex positions are adjusted to enforce
stealthy hyperuniformity of the full heterostructure as measured by the
spectral density $\tilde{\chi}_{\scriptscriptstyle V}(k)$ of this
two-phase medium. A gradient-based optimization is applied to minimize
the sum over $\tilde{\chi}_{\scriptscriptstyle V}(k)^{2}/k$ (for $k \leq
K_2$) by updating the vertex positions directly.

\begin{figure}[tbp]
\centering
\includegraphics[width=\textwidth]{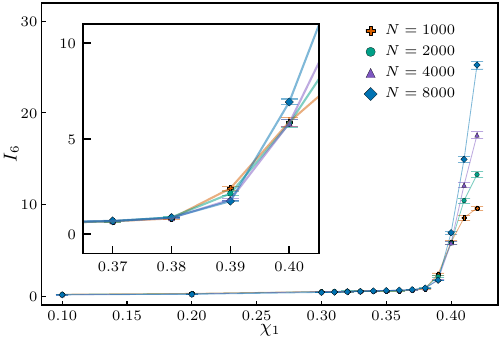}
\caption{Disorder-to-order phase transition identified by the
bond-orientational order parameter $I_6$, plotted as a function of
$\chi_{1}(K_1)$ for different system sizes (number of vertices $N$),
where each data point represents an average over 128 independent
realizations. The error bars denote one standard deviation. The inset
provides a detailed view of the critical region $\chi \in [0.37, 0.4]$.
The data exhibit clear evidence for a disorder-to-order phase transition
at $\chi \approx 0.39$.}
\label{fig:Psi6_correlation_integral_against_chi_progenitors}
\end{figure}

\section{Spectral characterization and network topology}

Our goal is to generate disordered stealthy hyperuniform trivalent
networks through controlled structural design, thereby generating short-
and long-range correlations while maintaining isotropy. To this end, we
choose the highest possible degree of stealthiness of the network vertices,
as quantified by $\chi_1(K_1)$, to establish a robust starting
configuration for the subsequent network refinement stage. To identify
the highest possible value of $\chi_1(K_1)$, we investigate the range of
$\chi_1(K_1)$ for which the structure factor can be minimized without
compromising the inherent disorder of the system. To measure the degree
of order in the system, we define a bond-orientational order parameter
$I_6$ that measures the decay of the correlation of bond-orientational
order $\psi_6$~\cite{steinhardt_bond-orientational_1983}; see method
section.  

In Figure \ref{fig:Psi6_correlation_integral_against_chi_progenitors},
we show the dependence of $I_6$ on $\chi_1(K_1)$, the control parameter
that determines the extent of $k$-space suppression in the structure
factor minimization. We compare multiple network sizes. We observe a
significant increase of $I_{6}$ at the critical value $\chi_{c} \approx
0.39$ that becomes sharper as the system size increases. This
observation is a clear indication of a phase transition from a
disordered to an ordered phase. A similar behavior is observed when
optimizing free point patterns in two dimensions towards stealthy
hyperuniformity, but the critical value of $\chi_{c} \approx 0.39$ is
significantly smaller in our case compared to the fully free point
pattern. This difference stems from the constraints imposed by the
associated triangle tessellation ($\chi_{c, \text{free}} \approx 0.5$).
In this study, $\chi=0.37$ is chosen, sufficiently away from the critical point 
to suppress finite-size effects and ensure isotropy and structural disorder of the networks. 
We achieve a significant suppression of the structure factor $S(k)$ by multiple orders of
magnitude (see SI). The same observation holds for the spectral density
of the networks generated after the first optimization stage before the
refinement. This suppression of single scattering makes them well suited
as start configurations for an optimization towards true stealthy
hyperuniformity of the networks in the second stage.

\begin{figure}[tbp]
\centering
\includegraphics[width=\columnwidth]{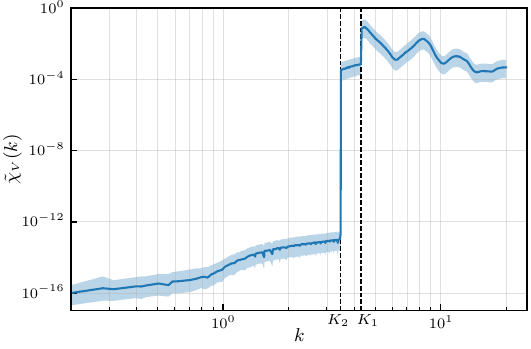}
\caption{Spectral density $\tilde{\chi}_{\scriptscriptstyle V}(k)$
of our stealthy hyperuniform networks (composed of disks and rectangles).
The solid line shows the median of $\tilde{\chi}_{\scriptscriptstyle
V}(k)$ for 500 realizations.
The shaded regions represent the interval between the 16th
and 84th percentiles of the distribution within each bin (with width $\Delta k
\approx 0.0199$).}
\label{fig:spectral_density_network_ensemble}
\end{figure}

In the second stage, we set $\chi_2(K_2) = 0.24$ as the stealthy
hyperuniformity parameter, corresponding to a range of wavevectors up to
$K_2$ where the spectral density $\tilde{\chi}_{\scriptscriptstyle
V}(k)$ is minimized. Choosing a larger $\chi_2(K_2)$ results in the
wavelength of $K_2$ approaching the typical size of the network's
structural cells, leading to significant deformation of the network
during optimization. These distorted configurations exhibit unfavorable
photonic properties, including the emergence of localized modes within
the band gap. It remains unclear whether these deformations for
$\mathbf{k}$-vectors with wavelength close to the network cell size
arise from our optimization protocol, such as the specific constraints
imposed on network topologies in the first stage, or represent an
intrinsic limitation of trivalent, disordered stealthy hyperuniform
networks. Therefore, the networks investigated in later parts of study
are characterized by the specific values $\chi_1(K_1) = 0.37$ (first
stage) and $\chi_2(K_2) = 0.24$ (second stage), which balance structural
stability with optimal scattering suppression. The radially averaged
spectral density of an ensemble of $N = 2000$-vertex networks is shown
in Figure~\ref{fig:spectral_density_network_ensemble}. The spectral
density are characterized by two distinct regimes: a region of true
stealthy hyperuniformity for $k \lesssim K_2$, and suppression of
$\tilde{\chi}_{\scriptscriptstyle V}(k)$ for multiple orders of
magnitude over multiple in the range $K_1 < k < K_2$, forming a ring in
$k$-space. This intermediate suppression is a direct consequence of the
initial topology optimization for $\chi_1(K_1) = 0.37$ and is essential
for the large, complete photonic band gaps observed in this work. This
structural feature is consistent with the network’s topology: as
$\chi_1(K_1)$ increases, the ring statistics of the networks improve,
with a higher fraction of hexagons and fewer other polygons (especially
squares and octagons). The networks predominantly consist of irregular
pentagons, hexagons, and heptagons, topologies known to minimize defect
states due to the established heuristics that the presence of squares
and octagons promotes localized modes and reduces therefore the band gap
size. Figure~\ref{fig:network_sample} shows a representative network
with $N = 2000$ vertices and parameters $\chi_1(K_1) = 0.37$,
$\chi_2(K_2) = 0.24$. The inset displays the spectral density,
confirming the isotropy of the ensemble through rotational symmetry in
$k$-space.

\section{Photonic properties of the networks}

The band structures of the networks are computed using a supercell
approximation, assuming periodic repetition of the structure in space,
and employing the highly accurate plane wave expansion method. Such
high-precision techniques are limited to 
networks containing with a few thousand vertices, preventing them from
capturing the true characteristics of large scale systems. 
%approaching the thermodynamic limit.

Consequently, evaluating only a single medium sized realization, as is
common in many prior studies, is insufficient to determine whether
photonic band gaps (PBGs) persist for large-scale applications or have
the possibility of remaining open in the thermodynamic limit. To
overcome this limitation, we adopt the established density of states
(DoS) ensemble method. By stacking the DoS from a large number of
modest-sized, independently generated realizations, we assess convergence as
the ensemble size increases. This approach enables to investigate
whether apparent band gaps close, to determine how robust the band gap
remain with increasing system size, and to characterize the shape of the
band tails.

We compare the DoS of networks generated with our novel method
($\chi_1(K_1) = 0.37$, $\chi_2(K_2) = 0.24$) against those from the
prior state-of-the-art approach, separately for TE- and TM-polarized
modes, as shown in Figure~\ref{fig:DOS_comparison}. This evaluation was
done for a dielectric contrast of 13.0 for both network types. We
performed a rough optimization of the disk radii ($r=0.24$) and wall
widths ($w=0.08$) for the stealthy hyperuniform networks by optimizing
the complete band gap for a few small networks (staying close to
established values for the networks generated by the method of
\cite{florescu_designer_2009}, $r=0.24$ and $w=0.10$). In addition, we
highlight in Figure~\ref{fig:DOS_comparison} the maximal complete band
gap, which corresponds to the largest frequency range devoid of states
in the DoS ensemble and calculate its gap-to-midgap ratio
$\Delta\omega/\omega_m$ where $\Delta\omega$ is the band gap width and
$\omega_m$ is the mid-frequency between the upper and lower band edge of
the ensemble. For our ensemble of 500 realizations, each containing 2000
vertices, we observe a complete PBG with a gap-to-midgap ratio of
$14.7\%$. Our stealthy hyperuniform networks support a significantly
larger complete photonic band gap compared to networks generated with
the previous best method, for which we obtain a complete PBG with a
gap-to-midgap ratio of $4.0\%$ for $r=0.24$ and $w=0.10$. When compared
to a honeycomb crystal (dielectric contrast 13.0) with identical network
parameters ($r = 0.24$, $w = 0.08$), the corresponding ratio is
$\Delta\omega/\omega_m \approx 16.7\%$. For network parameters
specifically optimized for this crystalline arrangements at the same
dielectric contrast ($r = 0.20$, $w = 0.046$), the gap-to-midgap ratio
reaches $\Delta\omega/\omega_m \approx 24.2\%$. Note that such a precise
optimization has not been carried out for our disordered network because
of the high computational cost (since at each step of the optimization
the band structures have to be computed for the entire ensemble).

The enhancement by our new inverse design method arises from an
unprecedented near-complete overlap between the TE- and TM-polarized
band gaps, combined with exceptionally sharp band edges, attributed to a
significant smaller fraction of localized defect states. The
localization of these edge states was investigated using the inverse
participation ratio (IPR) \cite{imagawa_photonic_2010,
dal_negro_waves_2021}, measuring the localization of the eigenstates,
and the results are provided in the Supplementary Information (SI). 

Additionally, we observe for the single best realization, a gap-to-midgap ratio 
$\Delta\omega/\omega_m \approx 19.0\%$;
see SI for its band structure and corresponding network configuration.
This gap-to-midgap ratio exceeds the best result obtained with the previous
method of $\Delta\omega/\omega_m = 12.6\%$ for the same ensemble size.
Hence, our approach also improves the band gap significantly already at small system sizes.
For an application using a single realization, the gap-to-midgap ratio can be further increased
by optimizing the disk radius and wall thickness.

\section{Discussion}

We identify two key findings from this work. The first one is an advance in 
fundamental physics:
showing that it is possible to have wide isotropic complete PBGs with
nearly perfect TE-TM overlap.
The second one is a breakthrough when it comes to practical applications:
identifying a pathway for photonic devices that take advantage of the isotropy of the resulting networks with hardly any cost in bandwidth compared to photonic crystals.

In terms of fundamental physics, 
we were able to use the ensemble approach to explore effective system sizes of up to 100,000 vertices
and observed wide complete PBGs whose band tails show no indication of closing.
These stealthy hyperuniform networks stand as the most promising candidate to date for an
isotropic two-dimensional amorphous material with a complete PBG in the
thermodynamic limit. Of course, a numerical study can probe this 
only up to some finite size.

When it comes to  applications, 
our heterostructures with unprecedented large complete PBGs can be implemented for most if not all  system sizes of practical interest.
The  structures proposed in this paper have the potential
to advance a range of future photonic technologies that require
a large system size, isotropy, and wide band gaps.
These technologies include free-form waveguides and photonic circuits, cavities with arbitrary orientation used for filters and sensors, and isotropic thermal absorbers and emitters.
The fabrication of
these photonic heterostructures is made feasible in recent years due to
advancements in high-precision additive manufacturing, e.g., 3D printing 
on a wide range of scales~\cite{ledermann_three-dimensional_2006,
wong_review_2012,
stefik_block_2015,
tumbleston_continuous_2015,
shirazi_review_2015,
bhushan_overview_2017,
uribe_experimental_2023-1,
zhang_printing_2024-1,
zhang_nanoscale_2025-1,
barsukova_stealthy-hyperuniform_2026}.
If the resolution of the fabrication method sets a lower limit
on the smallest feature in the photonic network, the wall width,
then the best rendition of our ensemble may even produce a wider band gap
than the honeycomb-based photonic crystal.

We expect that our current structures are not fully optimized.
First, the current parameters $r$ and $w$ are likely not optimal,
as our rough optimization does not fully resolve the fine-scale behavior near
the band edges and the optimal configuration also depends on the
specific dielectric contrast. 
Secondly, while our approach achieves significant suppression of the spectral density 
for wavenumbers $K_2< k < K_1$ (i.e., for wavelengths shorter than
the typical cell size), we plan to explore methods that may further suppress 
the spectral density in this regime while maintaining isotropy and 
statistical homogeneity of bond angles and lengths.
We anticipate that if such networks exist, they
will enable even slightly larger complete PBGs with steeper band edges.

\appendix

\section*{Methods}

\subsubsection*{\label{sec:order_parameter}Bond-orientational order parameter}
To measure the disorder, we look at the distance dependent
decay of the correlation of the local bond-orientational order
\cite{steinhardt_bond-orientational_1983} of the
progenitor points: 
\begin{equation}
g_{6}(r)=
\frac{
\displaystyle
\big\langle
\psi_{6}(i)\,\psi_{6}^{*}(j)\,
\delta\!\bigl(r-|\mathbf r_{i}-\mathbf r_{j}|\bigr)
\big\rangle_{i\neq j}
}{
\displaystyle
\big\langle
\delta\!\bigl(r-|\mathbf r_{i}-\mathbf r_{j}|\bigr)
\big\rangle_{i\neq j}
}\; .
\label{eq:g6_continuous}
\end{equation}
Here the angular brackets denote an average over all unordered pairs
$(i,j)$ and the local bond-orientational order parameter $\psi_6(j)$ to
track the disorder of the vertices of the triangulation and therefore
the formed centroid pattern that have an associated network structure: 
\begin{equation}
  \psi_6(j) = \frac{1}{|n_j|} \sum_{l \in n_j} e^{ i  6  \theta_{jl}},    
\end{equation}
where $\theta_{jl} =\arg(\vec{r}_l - \vec{r}_j)$ denotes the angle of
the vector from triangle vertex $j$ to $l$, measured in reference to a
global axis. 
%, and lies in $(-\pi, \pi]$.
A neighbor of a progenitor $j$ is another progenitor $l$ that shares an
edge with $j$ in the triangulation. To quantify how
rapidly the six‑fold bond‑orientational correlation $g_{6}(r)$ decays
with the distance, we define the integrated orientational order
$$I_{6}\;=\;\int_{0}^{L/2} g_{6}(r)\ \mathrm{d}r ,$$
where $L$ is the linear size of the unit cell. 

\subsubsection*{Network generation}

The full two-stage optimization framework is described in detail in the
SI. Here, we summarize the key network parameters and evaluation setup. 

The dependence of the bond orientational order parameter $I_6$ on the
stealthiness parameter $\chi_1$ is analyzed across network sizes $N \in
\{1000, 2000, 4000, 8000\}$, with 128 independent realizations per
$\chi_1$ evaluated over the interval $[0.1, 0.42]$. These centroid
patterns are generated using the first stage of our optimization
approach.

For the evaluation of photonic properties, we use 500 networks of $N
= 2000$ vertices generated via the complete two-stage procedure, with
$\chi_1(K_1) = 0.37$ and $\chi_2(K_2) = 0.24$. To benchmark the
performance of our method against the established approach of Florescu
\textit{et al.}~\cite{florescu_designer_2009}, we generate a comparative
dataset of 500 networks. These are derived from point patterns of 1,000
points exhibiting high-fidelity stealthy hyperuniformity ($\chi =
0.48$), which are then processed to yield networks of 2,000 vertices. In
this work, the number density is defined as the number of network
vertices per unit area, and we adopt units in which this density is set
to unity. 

\subsubsection*{Photonic density of states}

The photonic eigenvalues are computed at the $\Gamma$-point. Due to the
isotropy of the networks, this single $k$-point is sufficient to
determine the essential features of the band structure. The photonic
band structure is computed using the open-source software package
\textit{MIT Photonic Bands} (MPB)~\cite{johnson_block-iterative_2001-2}.
The MPB calculations requires a pixelation of the heterostructure which
degrades for $k<K_2$ the suppression of the spectral density to about
$10^{-7}$. We choose a resolution parameter of $16$ ($20$ for the
crystal structures), an eigensolver tolerance of $10^{-5}$, and a mesh
size used for smoothing the dielectric contrast of $5$. The DoS is
approximated using a histogram of the frequencies of the ensemble (each
ensemble contains 500 realizations) with a bin width of $\Delta \omega
\approx 0.00092$. The width of the complete PBG in the honeycomb crystal
was optimized using Brent's method, yielding a gap-to-midgap ratio of
$\Delta\omega/\omega_m \approx 24.2\%$. For the photonic crystals,
the band structure is evaluated along the edges of the irreducible
Brillouin zone using 40 values along each edge. 

\section*{Acknowledgements}

We thank Luca Lotz and Jaeuk Kim for valuable discussions.
This work was supported by the Initiative and Networking Fund of the
Helmholtz Association under the call Helmholtz Young Investigator Groups
(VH-NG-19-34, DataMat) and by the Deutsche Forschungsgemeinschaft (DFG,
German Research Foundation) through the SPP 2265, under grant number KL
3391/2-2.
P.J.S. and S.T. acknowledge funding and support of the U.S. Army Research
Office and was accomplished under Cooperative Agreement No. W911NF-22-2-0103.

\providecommand{\noopsort}[1]{}

\end{document}